\documentclass[twocolumn,showpacs,preprintnumbers,amsmath,amssymb,superscriptaddress]{revtex4}

\usepackage{graphicx}
\begin{document}
\bibliographystyle{prsty}
\title{Chemical potential shift and spectral weight transfer in Pr$_{1-x}$Ca$_x$MnO$_3$ revealed by photoemission spectroscopy}

\author{K. Ebata}
\affiliation{Department of Complexity Science and Engineering and Department of Physics, University of Tokyo, 
5-1-5 Kashiwanoha, Kashiwashi, Chiba, 277-8561, Japan}
\author{H. Wadati}
\affiliation{Department of Complexity Science and Engineering and Department of Physics, University of Tokyo, 
5-1-5 Kashiwanoha, Kashiwashi, Chiba, 277-8561, Japan}
\author{M. Takizawa}
\affiliation{Department of Complexity Science and Engineering and Department of Physics, University of Tokyo, 
5-1-5 Kashiwanoha, Kashiwashi, Chiba, 277-8561, Japan}
\author{A. Fujimori}
\affiliation{Department of Complexity Science and Engineering and Department of Physics, University of Tokyo, 
5-1-5 Kashiwanoha, Kashiwashi, Chiba, 277-8561, Japan}
\author{A. Chikamatsu}
\affiliation{Department of Applied Chemistry, University of Tokyo, 
Bunkyo-ku, Tokyo 113-8656, Japan}
\author{H. Kumigashira}
\affiliation{Department of Applied Chemistry, University of Tokyo, 
Bunkyo-ku, Tokyo 113-8656, Japan}
\author{M. Oshima}
\affiliation{Department of Applied Chemistry, University of Tokyo, 
Bunkyo-ku, Tokyo 113-8656, Japan}
\author{Y. Tomioka}
\affiliation{Correlated Electron Research Center (CERC), National Institute of Advanced Industrial Science and Technology (AIST), Tsukuba 305-8562, Japan}
\author{Y. Tokura}
\affiliation{Department of Applied Physics, University of Tokyo Bunkyo-ku, Tokyo 113-8656, Japan}
\affiliation{Correlated Electron Research Center (CERC), National Institute of Advanced Industrial Science and Technology (AIST), Tsukuba 305-8562, Japan}
\affiliation{Spin Superstructure Project, Exploratory Research for Advanced Technology (ERATO), Japan Science and Technology Corporation (JST), Tsukuba 305-8562, Japan}
\date{\today}

\begin{abstract}
We have studied the chemical potential shift and changes in the electronic density of states near the Fermi level ($E_F$) as a function of carrier concentration in Pr$_{1-x}$Ca$_x$MnO$_3$ (PCMO, $0.2 \le x \le 0.65$) through the measurements of photoemission spectra. The results showed that the chemical potential shift was suppressed for $x \agt 0.3$, where the charge exchange (CE)-type antiferromagnetic charge-ordered state appears at low temperatures. We consider this observation to be related to charge self-organization such as stripe formation on a microscopic scale in this composition range. Together with the previous observation of monotonous chemical potential shift in La$_{1-x}$Sr$_x$MnO$_3$, we conclude that the tendency toward the charge self-organization increases with decreasing bandwidth. In the valence band, spectral weight of the Mn 3$d$ $e_g$ electrons in PCMO was transferred from $\sim$ 1 eV below $E_F$ to the region near $E_F$ with hole doping, leading to a finite intensity at $E_F$ even in the paramagnetic insulating phase for $x \agt 0.3$, probably related with the tendency toward charge self-organization. The finite intensity at $E_F$ in spite of the insulating transport behavior is consistent with fluctuations involving ferromagnetic metallic states.
\end{abstract}


\maketitle
\section{Introduction}
In recent decades, manganites with chemical compositions $R_{1-x}A_x$MnO$_3$, where $R$ is a rare-earth ($R$=La, Nd, Pr) and $A$ an alkaline-earth metal ($A$=Sr, Ba, Ca), have been extensively studied because of their remarkable physical properties such as colossal magnetoresistance (CMR) and spin, charge and orbital ordering \cite{Tokura}. Historically, their magnetic and transport properties have been discussed in terms of double-exchange (DE) mechanism \cite{Zener, Anderson}. Half-doped manganites ($x$ $\simeq$ 0.5) have been a focus of recent studies because most of them exhibit a so-called charge exchange (CE)-type antiferromagnetic (AFM) charge-ordered (CO) state. The compound Pr$_{1-x}$Ca$_x$MnO$_3$ (PCMO), in which the bandwidth $W$ is small in comparison with other manganites, has a particularly stable CO state over the wide hole concentration region between $x$ $\simeq$ 0.3 and 0.75 as shown in the electronic phase diagram in Fig. 1 \cite{Zimmermann2}. Furthermore, CMR in the CO state of PCMO has been found remarkable, amounting to several orders of magnitude \cite{tomioka}.
\begin{figure}
\begin{center}
\includegraphics[width=6cm]{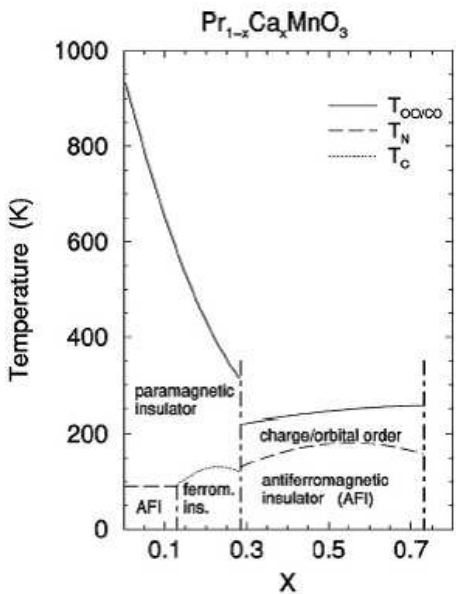}
\caption{Electronic phase diagram of Pr$_{1-x}$Ca$_x$MnO$_3$ \cite{Zimmermann2}.}
\label{phase}
\end{center}
\end{figure}
The origin of the CMR effect has been debated in recent years since it was pointed out that DE mechanism alone was insufficient to explain the large resistivity change with magnetization \cite{Millis}. Recently, the possibility of charge self-organization such as phase separation (PS) as the origin of the CMR effect has been discussed in theoretical and experimental studies. In computational work based on quantum Monte Carlo simulation, the possibility of PS into the ferromagnetic (FM) and AFM phases has been pointed out \cite{Yunoki, Moreo1}. The calculated density of electrons $\langle n \rangle$ versus chemical potential $\mu$ for a one-dimensional model has shown a clear discontinuity in $\langle n \rangle$ at a particular value of $\mu$, indicating a PS \cite{Yunoki, Moreo1} since in general electronic PS results in the pinning of $\mu$. Indeed, electron microscopy studies have revealed inhomogeneous spatial images in which the FM charge-disordered and AFM CO microdomains coexist \cite{Mori1, Uehara, Sarma}. However, it has been unclear whether they are driven by (intrinsic) tendency toward electronic PS or triggered by (extrinsic) chemical inhomogenity. Electronic ``PS" on a nano-meter scale such as stripe formation would also lead to the pinning of the chemical potential, in analogy to the case of high-$T_c$ cuprates \cite{Ino}. It has also been shown theoretically that the introduction of chemical disorder turns the chemical potential pinning into a smooth shift \cite{Moreo2}. Therefore, one can distinguish intrinsic electronic inhomogenity from extrinsic chemical disorder through the measurements of chemical potential shift. 

The chemical potential shift can be deduced from the core-level shifts of photoemission spectra as functions of carrier concentration. Recently, it has been shown that the chemical potential shift in La$_{1-x}$Sr$_x$MnO$_3$ (LSMO), which has the widest bandwidth among the manganites, shows a monotonous shift without indication of chemical potential pinning, especially around $x$ $\simeq$ 0.3, where the CMR effect is the strongest \cite{Matsuno}. It has therefore been considered that there is no intrinsic electronic ``PS" in LSMO.
In spite of the smooth shift in the core levels, changes in the electronic structure of LSMO near the Fermi level ($E_F$) with carrier doping were highly non-rigid-band like in the sense that the Mn 3$d$ $e_g$ peak lost its intensity with hole doping while its position relative to $E_F$ remained nearly unchanged \cite{Saitoh, Horiba}.

In this paper, we report on a core-level photoemission study of the chemical potential shift $\Delta \mu$ as a function of carrier concentration and the composition dependence of the valence-band spectra in single crystals of PCMO. The results show that the chemical potential pinning occurs for $x \agt 0.3$, spectral weight is transferred toward near $E_F$ with hole doping, and there is a finite intensity at $E_F$ in the paramagnetic insulating (PI) phase on the high temperature side of the CO transition temperature for $x \agt 0.3$.

\section{Experimental}
Single crystals of PCMO with the carrier concentrations of $x=0.2$, 0.25, 0.3, 0.35, 0.45, 0.5, and 0.65 were grown by the floating-zone method. The growth techniques and transport properties of the crystals were described in ref.\cite{tomioka}. The photoemission spectroscopy (PES) measurements using synchrotron radiation were performed at BL-2C of Photon Factory, High Energy Acceleration Research Organization (KEK). We carried out the measurements using the photon energies of $h\nu =$ 600 eV, 643.6 eV (Mn 2$p$-3$d$ resonance), 800 eV, and 930 eV (Pr 3$d$-4$f$ resonance). Ultraviolet photoemission spectroscopy (UPS) and X-ray photoemission spectroscopy (XPS) measurements were also performed using the photon energies of $h\nu =$ 21.2 eV (He I) and 1253.6 eV (Mg $K \alpha$). All the photoemission measurements were performed under the base pressure of $\sim 10^{-10}$ Torr at room temperature. The samples were repeatedly scraped {\it in situ} with a diamond file to obtain clean surfaces. In the case of PES using synchrotron radiation and XPS, the cleanliness of the sample surface was checked by the reduction of the shoulder on the high binding energy side of the O 1$s$ core level. In the case of UPS, the scraping was made until a bump around 9-10 eV, which is attributed to surface contamination, decreased and the valence-band spectra did not change with further scrapings. Photoelectrons were collected using a Scienta SES-100 electron-energy analyzer. The energy resolution was about 200-500 meV for the synchrotron radiation measurements, 20 meV for He I, and 800 meV for Mg $K \alpha$. The measured binding energies were stable, judged from the fact the gold $4{\it f}_{7/2}$ core-level spectrum was nearly unchanged throughout the measurements.

\section{Results and discussion}
\subsection{Chemical potential shift}
In order to deduce the chemical potential shift from a set of core-level data, we utilize the formula \cite{Hufner, Fujimori} that the shift $\Delta E_B$ of the binding energy is given by $\Delta E_B$ = $\Delta \mu$ $+$ $K \Delta Q$ $+$ $\Delta V_M$ $-$ $\Delta E_R$, where $\Delta \mu$ is the change in the chemical potential, $K$ is the coupling constant of the Coulomb interaction between the valence and the core electrons, $ \Delta Q$ is the change in the number of valence electrons on the atom considered, $\Delta V_M$ is the change in the Madelung potential, and $\Delta E_R$ is the change in the extra-atomic relaxation energy. Here, $ \Delta Q$ produces changes in the electrostatic potential at the core-hole site as well as in the intra-atomic relaxation energy of the core-hole final state. $\Delta E_R$ is due to changes in the screening of the core-hole potential by metallic-conduction electrons and polarizable surrounding ions.

Figure 2 shows the photoemission spectra of the O $1s$, Ca $2p$, Pr $4d$, and Mn $2p$ core levels taken at $h\nu =$ 800 eV. The line shapes of most of the core levels did not change significantly for these compositions. Nevertheless, because the line shape on the higher binding energy side of the O $1s$ was sensitive to a slight surface degradation or contamination, we used the midpoint of the lower binding energy slope rather than the peak position for the O $1s$ core level to deduce the amount of the core-level shift more reliably. We also used the midpoint for Ca $2p$. As for Pr $4d$ and Mn 2$p$, the peak position was used because the line shape slightly changed with doping and therefore the midpoint position was less meaningful. Also, the photoemission spectra for $x=0.25$, 0.3, and 0.35 were measured at $h\nu =$ 1253.6 eV, and the shifts of $x=0.25$, and 0.35 relative to $x=0.3$ were determined.

\begin{figure}
\begin{center}
\includegraphics[width=8cm]{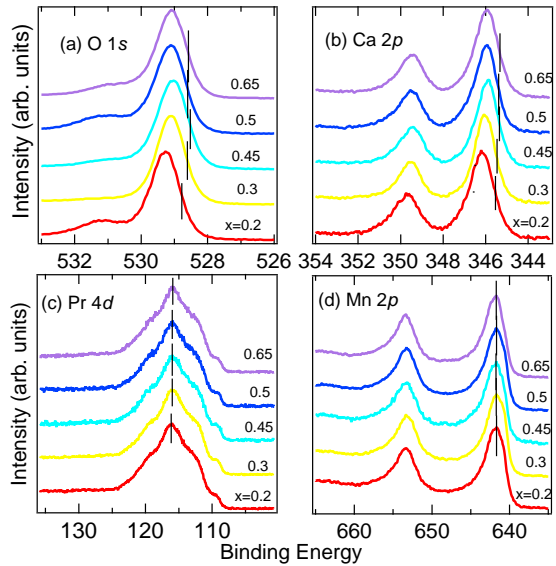}
\caption{(Color online) Core-level photoemission spectra of Pr$_{1-x}$Ca$_x$MnO$_3$ taken at $h\nu =$ 800 eV. (a) O $1s$; (b) Ca $2p$; (c) Pr 4$d$; (d) Mn 2$p$. The intensity has been normalized to the peak height.}
\label{core1}
\end{center}
\end{figure}

In Fig. 3(a), the binding energy shifts of the O $1s$, Ca $2p$, Pr $4d$, and Mn $2p$ core levels are plotted. One can see from Fig. 3(a) that the observed binding energy shifts with $x$ were approximately common to the O $1s$, Ca $2p$, and Pr $4d$ core levels, whereas the shift of the Mn $2p$ core level was different from them. The similar shifts of the O $1s$, Ca $2p$, and Pr $4d$ core levels indicate that the change in the Madelung potential $\Delta V_M$ has negligible effects on the core-level shifts because it would cause shifts of the core levels of the O$^{2-}$ anion and the Ca$^{2+}$ and Pr$^{3+}$ cations in different directions. Core-hole screening $\Delta E_R$ by conduction electrons can also be excluded from the main origin of the core-level shifts in transition-metal oxides \cite{Ino, Harima}. Therefore, we assume that $\Delta E_B$ $\simeq$ $\Delta \mu$ + $K \Delta Q$. The $K \Delta Q$ term is important for the Mn $2p$ core-level shift because of the increase in the Mn valence with hole doping $(\propto -\Delta Q)$ from Mn$^{3+}$ towards Mn$^{4+}$, ${i.e.,}$ because of the so-called chemical shift. Therefore, we consider that the shifts of the O $1s$, Ca $2p$, and Pr $4d$ core levels are largely due to the chemical potential shift $\Delta \mu$, and take the average of the shifts of the three core levels as a measure of $\Delta \mu$ in PCMO.

In Fig. 3(b), we have plotted $\Delta \mu$ in PCMO thus deduced as a function of carrier concentration. The shift was large in the region $x \alt 0.25$, which may be understood in analogy to the case of LSMO \cite{Matsuno}. The suppression of the chemical potential shift was observed near and in the CO region $x \agt 0.3$. If the suppression of the chemical potential shift is due to an electronic PS as the thermodynamic relationship suggests, in order to gain the long-range Coulomb energy the PS should occur only on a microscopic scale as in the bi-stripe or Wigner-crystal model suggested for La$_{1-x}$Ca$_x$MnO$_3$ for $x \geq 0.5$ \cite{Mori2, Radaelli, stripe}. We note that in La$_{2-x}$Sr$_x$CuO$_4$ and La$_{2-x}$Sr$_x$NiO$_4$ the suppression of the chemical potential shift is most likely related to the formation of stripes \cite{Ino, Satake}. Although our measurements were done in the PI phase above the CO transition temperature, fluctuations of the CO state are expected to be present in the PI phase as was observed by means of x-ray resonant scattering \cite{Zimmermann}. Also, from the observed suppression of $\Delta \mu$ in the region around $x \sim 0.3$, which exhibits the ferromagnetic insulating phase at low temperatures, we speculate that the fluctuation of the CO state also exist over a finite concentration range around the phase boundary at $x$ = 0.3.

Since the effect of disorder was found to strongly influence the chemical potential shift \cite{Moreo2}, disorder strength in PCMO has to be evaluated. According to the variance of the ionic radii of the A-site cation, the effect of disorder in PCMO as well as that in LSMO is very small \cite{Tomioka2}.
\begin{figure}
\begin{center}
\includegraphics[width=8.5cm]{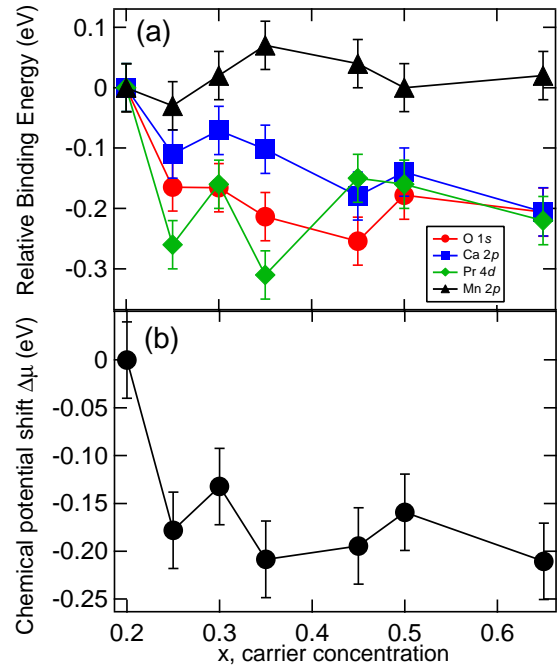}
\caption{(Color online) Core-level shifts and chemical potential shift. (a) Binding energy shifts of the O $1s$, Ca $2p$, Pr $4d$, and Mn $2p$ core levels as functions of carrier concentration $x$; (b) Chemical potential shift $\Delta \mu$ in Pr$_{1-x}$Ca$_x$MnO$_3$ as a function of carrier concentration $x$.}
\label{chemicalpotential}
\end{center}
\end{figure}
In Fig. 4, we compare $\Delta \mu$ for PCMO with that for LSMO. For $x \alt 0.25$, both $\Delta \mu$ curves show similar doping dependences in the sense that the shifts are monotonous. The difference between the two become pronounced for $x \agt 0.3$, where the CO state occurs or the fluctuation of CO state becomes significant in PCMO but not in LSMO. Therefore, the different chemical potential shifts in PCMO and LSMO indeed reflect the CO in PCMO, and the $\mu$ pinning in the CO state of PCMO can be understood as due to the stronger tendencies toward ``PS" in PCMO. Considering the suppression of $\Delta \mu$ in PCMO, the absence of such suppression in LSMO and the much stronger CMR in PCMO than in LSMO, it is possible that the tendency towards electronic ``PS" on a microscopic scale is the origin of CMR in manganite with CO state. More systematic studies on other manganites as well as temperature-dependent studies of the chemical potential shift will shed further light on the relationship between CMR and electronic ``PS".
\begin{figure}
\begin{center}
\includegraphics[width=8.5cm]{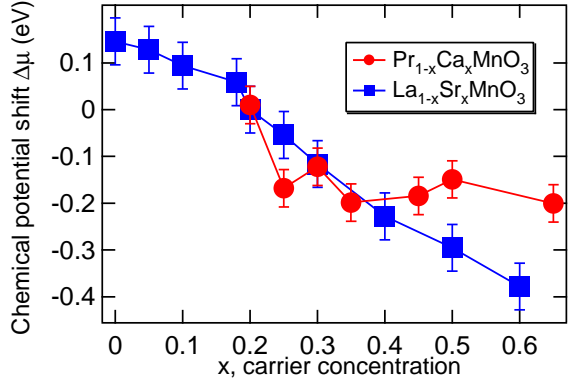}
\caption{(Color online) Comparison of the chemical potential shift in Pr$_{1-x}$Ca$_x$MnO$_3$ with that in La$_{1-x}$Sr$_x$MnO$_3$ \cite{Matsuno}.}
\label{LSMOshift}
\end{center}
\end{figure}

\subsection{Spectral weight transfer near $E_F$}

Valence-band spectra of PCMO taken at different photon energies are shown in Fig. 5(a). The spectra mainly consisted of structures as labeled A, (A',) B, C, (C', C",) and D. The spectra taken at $h\nu =$ 930 and 643.6 eV correspond to Pr 3$d$-4$f$ and Mn 2$p$-3$d$ on-resonance PES spactra, respectively. From the Mn 2$p$-3$d$ resonance spectra, Mn 3$d$-derived features appeared as structures A', C', and D. Structures A' and C' are shifted slightly toward higher binding energies than A and C, respectively, due to the different matrix elements between normal PES and resonant PES. Also, the intensity of structure C" was strongly enhanced in the Pr 3$d$-4$f$ resonance spectra. The UPS ($h\nu =$ 21.2 eV) spectrum represents the O 2$p$ state due to the large relative photo-ionization cross-section of O 2$p$ at low photon energies. Therefore, structures A, B, C, and D in the UPS spectrum are assigned to the Mn 3$d$-O 2$p$ bonding state, the non-bonding O 2$p$ state, the Mn 3$d$ $t_{2g}$ plus the Pr 4$f$ states, and the Mn 3$d$ $e_g$ state, respectively, consistent with the cluster-model calculation for LSMO \cite{Saitoh}.
\begin{figure}
\begin{center}
\includegraphics[width=6.5cm]{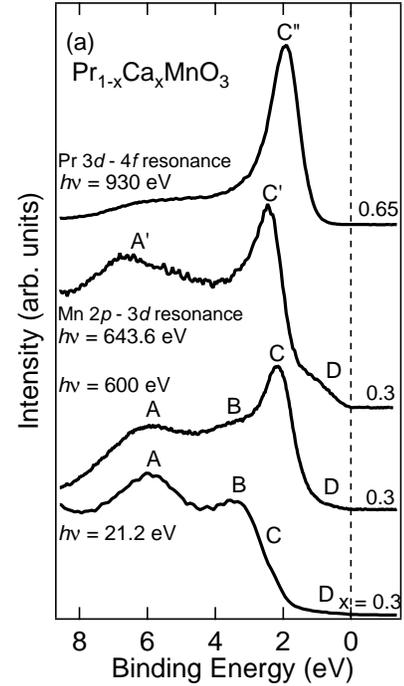}
\includegraphics[width=8cm]{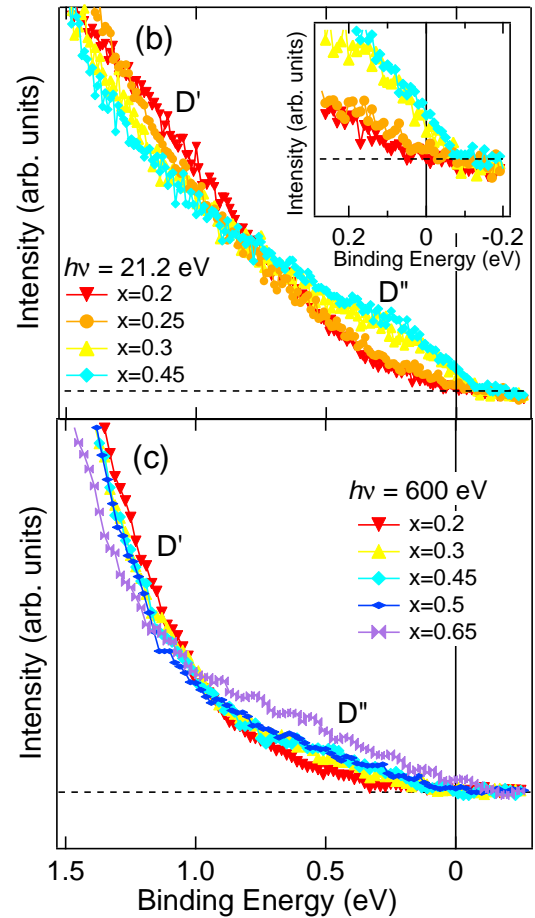}
\caption{(Color online) Valence-band photoemission spectra of Pr$_{1-x}$Ca$_x$MnO$_3$. (a) Comparison of spectra taken at various photon energies; (b) Spectra near $E_F$ taken at $h\nu =$ 21.2 eV; (c) $h\nu =$ 600 eV.}
\label{valence}
\end{center}
\end{figure}

Figure 5(b), 5(c) show the valence-band spectra near $E_F$ for various hole concentrations taken at $h\nu =$ 21.2 and 600 eV, respectively. The spectra have been normalized to the integrated intensity in the energy range from 1.5 eV to $\sim$ $E_F$. Two features labeled D' and D" were observed in these spectra. Spectral weight was transferred from D' to D" with increasing hole concentration, that is, the valence-band spectra near $E_F$ exhibited highly non-rigid-band-like behavior, similar to that in the LSMO \cite{Saitoh, Horiba}. Similar behaviors have been widely observed in filling-controlled transition-metal oxides \cite{Maiti}. This spectral weight transfer may be related to dynamical stripe formation because the hole-rich and hole-poor regions would give rise to the low and high binding energy features, respectively, and the relative contributions of the hole-rich regions would increase with hole doping \cite{Ino2, Yoshida}.

Next, we discuss the origin of the finite spectral weight at $E_F$ in the insulating phase of PCMO. The spectra in a narrow energy region near $E_F$ are shown in the inset of Fig. 5(b). For $x$ = 0.2, and 0.25, the insulating gap was opened. For $x \agt 0.3$, a finite intensity at $E_F$ was observed in the PI phase, leading to the pseudogap behavior. Recently, a finite intensity at $E_F$ was observed in the photoemission spectra of Nd$_{1-x}$Sr$_x$MnO$_3$ in the PI phase above the Curie temperature of the FM metallic phase \cite{Sekiyama}.
Kajimoto {\it et al}. \cite{Kajimoto} have remarked that FM fluctuations exist in the PI phase of the PCMO by means of neutron scattering studies. Theoretically, FM fluctuations in the CO state of PCMO was found to lead to metallic character using a combination of numerical relaxation technique and analytic mean-field approximation \cite{Hotta}. On the other hand, from the x-ray resonant scattering studies, the CO fluctuations were also observed in the PI phase of the PCMO \cite{Zimmermann}. Therefore, we consider that the pseudogap features for $x \agt 0.3$, namely, the density of states (DOS) minimum and the finite intensity at $E_F$ are caused by the CO fluctuations and FM fluctuations, respectively. Monte Carlo simulations on the PS model have shown that the pseudogap behavior is caused by the formation of FM metallic clusters in an insulating background, and the resultant mixed-phase state is crucial for the pseudogap formation \cite{Moreo3, Aliaga}. In the PI phase of Bi$_{1-x}$Ca$_x$MnO$_3$, an atomic-scale image of PS into FM metallic and insulating regions was observed by means of scanning tunneling microscopy, and it is considered that the FM fluctuations are related to thermally activated hopping of $e_g$ electrons induced by DE mechanism at high temperatures \cite{Renner, Bao, Liu}. 
Recently, by angle-resolved photoemission measurements, a pseudogap was observed in the low-temperature FM state of the bilayer compound La$_{1.2}$Sr$_{1.8}$Mn$_2$O$_7$ \cite{Dessau, Chuang}. The observations of the pseudogap behavior in manganites may provide an important clue for understanding the origin of the CMR effect.

\section{Conclusion}
We have experimentally determined the chemical potential shift $\Delta \mu$ as a function of carrier concentration in PCMO by means of core-level photoemission spectroscopy and observed a suppression of the shift near and in the CO composition range. This result indicates that there is charge self-organization such as electronic ``PS" on a microscopic scale which has been suggested by the bi-stripe or Wigner-crystal model for La$_{1-x}$Ca$_x$MnO$_3$ \cite{Mori2, Radaelli}. Comparison with the chemical potential shift in LSMO implies that the electronic ``PS" may be related to the origin of CMR in manganites with CO state. In the valence band near $E_F$ of PCMO, spectral weight was transferred with hole doping, and for the CO region $x \agt 0.3$ we observed the pseudogap behavior with a finite DOS at $E_F$, consistent with the presence of both FM and CO fluctuations in the PI phase.

\section{Acknowledgment}
Informative discussion with S.-W. Cheong is gratefully acknowledged. This work was supported by a Grant-in-Aid for Scientific Research in Priority Area ``Invention of Anomalous Quantum Materials" from the Ministry of Education, Culture, Sports, Science and Technology, Japan. This work was done under the approval of the Photon Factory Program Advisory Committee (Proposal No. 2005G101).


\begin{thebibliography}{10}

\bibitem{Tokura}
Y. Tokura and N. Nagaosa, Science {\bf 288},  462  (2000).

\bibitem{Zener}
C. Zener, Phys. Rev. {\bf 82},  403  (1951).

\bibitem{Anderson}
P.~W. Anderson and H. Hasegawa, Phys. Rev. {\bf 100},  675  (1955).

\bibitem{Zimmermann2}
M. v.~Zimmermann, C.~S. Nelson, J.~P. Hill, D. Gibbs, M. Blume, D. Casa, B.
  Keimer, Y. Murakami, C.-C. Kao, C. Venkataraman, T. Gog, Y. Tomioka, and Y.
  Tokura, Phys. Rev. B {\bf 64},  195133  (2001).

\bibitem{tomioka}
Y. Tomioka, A. Asamitsu, H. Kuwahara, Y. Moritomo, and Y. Tokura, Phys. Rev. B
  {\bf 53},  R1689  (1996).

\bibitem{Millis}
A.~J. Millis, P.~B. Littlewood, and B.~I. Shraiman, Phys. Rev. Lett. {\bf 74},
  5144  (1995).

\bibitem{Yunoki}
S. Yunoki, J. Hu, A.~L. Malvezzi, A. Moreo, N. Furukawa, and E. Dagotto, Phys.
  Rev. Lett. {\bf 80},  845  (1998).

\bibitem{Moreo1}
A. Moreo, S. Yunoki, and E. Dagotto, Science {\bf 283},  2034  (1999).

\bibitem{Mori1}
S. Mori, C.~H. Chen, and S.-W. Cheong, Phys. Rev. Lett. {\bf 81},  3972
  (1998).

\bibitem{Uehara}
M. Uehara, S. Mori, C.~H. Chen, and S.-W. Cheong, Nature {\bf 399},  560
  (1999).

\bibitem{Sarma}
D.~D. Sarma, D. Topwal, U. Manju, S.~R. Krishnakumar, M. Bertolo, S.~La Rosa,
  G. Cautero, T.~Y. Koo, P.~A. Sharma, S.-W. Cheong, and A. Fujimori, Phys.
  Rev. Lett. {\bf 93},  097202  (2004).

\bibitem{Ino}
A. Ino, T. Mizokawa, A. Fujimori, K. Tamasaku, H. Eisaki, S. Uchida, T. Kimura,
  T. Sasagawa, and K. Kishio, Phys. Rev. Lett. {\bf 79},  2101  (1997).

\bibitem{Moreo2}
A. Moreo, M. Mayr, A. Feiguin, S. Yunoki, and E. Dagotto, Phys. Rev. Lett. {\bf
  84},  5568  (2000).

\bibitem{Matsuno}
J. Matsuno, A. Fujimori, Y. Takeda, and M. Takano, Europhys. Lett. {\bf 59},
  252  (2002).

\bibitem{Saitoh}
T. Saitoh, A.~E. Bocquet, T. Mizokawa, H. Namatame, A. Fujimori, M. Abbate, Y.
  Takeda, and M. Takano, Phys. Rev. B {\bf 51},  13942  (1995).

\bibitem{Horiba}
K. Horiba, A. Chikamatsu, H. Kumigashira, M. Oshima, N. Nakagawa, M. Lippmaa,
  K. Ono, M. Kawasaki, and H. Koinuma, Phys. Rev. B {\bf 71},  155420  (2005).

\bibitem{Hufner}
S. H\"{u}fner, {\em Photoelectron Spectroscopy} (Springer-Verlag, Berlin, 2003).

\bibitem{Fujimori}
A. Fujimori, A. Ino, J. Matsuno, T. Yoshida, K. Tanaka, and T. Mizokawa, J.
  Electron Spectrosc. Relat. Phenom. {\bf 124},  127  (2002).

\bibitem{Harima}
N. Harima, J. Matsuno, A. Fujimori, Y. Onose, Y. Taguchi, and Y. Tokura, Phys.
  Rev. B {\bf 64},  220507(R) (2001).

\bibitem{Mori2}
S. Mori, C.~H. Chen, and S.-W. Cheong, Nature {\bf 392},  473  (1998).

\bibitem{Radaelli}
P.~G. Radaelli, D.~E. Cox, L. Capogna, S.-W. Cheong, and M. Marezio, Phys. Rev.
  B {\bf 59},  14440  (1999).

\bibitem{stripe}
While Tokunaga {\it et al}. [J. Magn. Magn. Mater {\bf 226-230},  851  (2001)] observed a macroscopic PS with length scale exceeding one micrometer in PCMO at $x=0.3$ by means of magneto-optical measurements, the electronic ``PS" such as stripes will be driven on a microscopic scale.

\bibitem{Satake}
M. Satake, K. Kobayashi, T. Mizokawa, A. Fujimori, T. Tanabe, T. Katsufuji, and
  Y. Tokura, Phys. Rev. B {\bf 61},  15515  (2000).

\bibitem{Zimmermann}
M. v.~Zimmermann, J.~P. Hill, D. Gibbs, M. Blume, D. Casa, B. Keimer, Y.
  Murakami, Y. Tomioka, and Y. Tokura, Phys. Rev. Lett. {\bf 83},  4872
  (2003).

\bibitem{Tomioka2}
Y. Tomioka and Y. Tokura, Phys. Rev. B {\bf 70},  014432  (2004).

\bibitem{Maiti}
K. Maiti and D.~D. Sarma, Phys. Rev. B {\bf 61},  2525  (2000).

\bibitem{Ino2}
A. Ino, C. Kim, M. Nakamura, T. Yoshida, T. Mizokawa, Z.-X. Shen, A. Fujimori,
  T. Kakeshita, H. Eisaki, and S. Uchida, Phys. Rev. B {\bf 62},  4137  (2000).

\bibitem{Yoshida}
T. Yoshida, X. J. Zhou, T. Sasagawa, W. L. Yang, P. V. Bogdanov, A. Lanzara, Z. Hussain, T.
  Mizokawa, A. Fujimori, H. Eisaki, Z.-X. Shen, T. Kakeshita, and S. Uchida,
  Phys. Rev. Lett. {\bf 91},  027001  (2003).

\bibitem{Sekiyama}
A. Sekiyama, H. Fujiwara, A. Higashiya, S. Imada, H. Kuwahara, Y. Tokura, and
  S. Suga, cond-mat  0401601  (2004).

\bibitem{Kajimoto}
R. Kajimoto, T. Kakeshita, Y. Oohara, H. Yoshizawa, Y. Tomioka, and Y. Tokura,
  Phys. Rev. B {\bf 58},  R11837  (2003).

\bibitem{Hotta}
T. Hotta and E. Dagotto, Phys. Rev. B {\bf 61},  R11879  (2000).

\bibitem{Moreo3}
A. Moreo, S. Yunoki, and E. Dagotto, Phys. Rev. Lett. {\bf 83},  2773  (1999).

\bibitem{Aliaga}
H. Aliaga, D. Magnoux, A. Moreo, D. Poilblanc, S. Yunoki, and E. Dagotto, Phys.
  Rev. B {\bf 68},  104405  (2003).

\bibitem{Renner}
C. Renner, G. Aeppli, B.-G. Kim, Y.-A. Soh, and S.-W. Cheong, Nature {\bf 416},
   518  (2002).

\bibitem{Bao}
W. Bao, J.~D. Axe, C.~H. Chen, and S.-W. Cheong, Phys. Rev. Lett. {\bf 78},  543
  (1997).

\bibitem{Liu}
H.~L. Liu, S.~L. Cooper, and S.-W. Cheong, Phys. Rev. Lett. {\bf 81},  4684
  (1998).

\bibitem{Dessau}
D.~S. Dessau, T. Saitoh, C.-H. Park, Z.-X. Shen, P. Villella, N. Hamada, Y.
  Moritomo, and Y. Tokura, Phys. Rev. Lett. {\bf 81},  192  (1998).

\bibitem{Chuang}
Y.-D. Chuang, A.~D. Gromko, D.~S. Dessau, T. Kimura, and Y. Tokura, Science
  {\bf 292},  1509  (2001).

\end{thebibliography}
\end{document}